\documentstyle[prc,aps,preprint,eqsecnum,epsf]{revtex}
\tighten
\begin{document}

\draft
\title{Derivation and assessment of strong coupling core-particle 
model from the Kerman-Klein-D\"onau-Frauendorf theory}
\author{Pavlos Protopapas
\thanks{pavlos@walet.physics.upenn.edu}
and Abraham Klein
\thanks{aklein@walet.physics.upenn.edu}} 
\address{Department of Physics, University of Pennsylvania, Philadelphia,
PA 19104-6396}  

\date{\today}

\maketitle

\begin{abstract}
We review briefly the fundamental equations of a semi-microscopic
core-particle coupling method that makes no reference to an intrinsic
system of coordinates.  We then demonstrate how an intrinsic system
can be introduced in the strong coupling limit so as to yield a
completely equivalent formulation.  It is emphasized that the
conventional core-particle coupling calculation introduces a further
approximation that avoids what has hitherto been the most
time-consuming feature of the full theory, and that this approximation
can be introduced either in the intrinsic system, the usual case, or
in the laboratory system, our preference. A new algorithm is described
for the full theory that largely removes the difference in
complexity between the two types of calculation.  Comparison of the
full and approximate theories for some representative cases provides a
basis for the assessment of the accuracy of the traditional approach.
We find that for  well-deformed nuclei, 
e.g. $^{157}$Gd and $^{157}$Tb, the core-coupling method and the full theory
give similar results.
\end{abstract}

\begin{center}
PACS number(s): 21.60.-n, 21.60.Ev, 21.10.-k, 21.10.Re
\end{center}

\section{Introduction}

We have recently undertaken the task of revitalizing and extending 
a semi-microscopic
theory of collective motion for odd nuclei that we shall refer to as the
Kerman-Klein-D\"onau-Frauendorf (KKDF) model \cite{pav:1,pav:2,pav:3,pav:4}.
This model, aside from the elements discussed for the first time in the
present paper, was introduced in close to its present form by D\"onau and
Frauendorf \cite{DF1,DF2,DF3,DF4,DF5}, whose work was in turn stimulated 
by an application
\cite{Dreiss} of the theory of collective motion developed by Kerman and 
Klein \cite{kk1,kk2,kk3,kk4}.

In the presentation of our work at seminars and conferences, 
one question that has invariably
arisen is the connection between the KKDF model and the conventional
core-particle coupling model, especially for deformed nuclei, to which
our published applications have so far been confined.  Even if we widen
the inquiry to the connection between the shell model and the core-particle
model, we find that the literature on this subject is
sparse.  We are aware of only two publications that have been addressed
specifically to this topic. The earlier of these papers \cite{cpc} showed how
all, then extant, core-particle coupling models could be understood as
approximations to the work of Kerman and Klein.  This paper appears 
to have gone completely unnoticed, since it is not quoted in the later work
\cite{Brink}, which is devoted to the derivation of the strong coupling
core-particle model from a schematic shell model.  In the book by Ring
and Schuck \cite{RS}, which appeared betweentimes, 
the success of the strong coupling model in its domain
of application is heralded but at the same time proclaimed a mystery.  

The main purposes of the present work are threefold.  
The first is to transform
the Kerman-Klein equations from the ``laboratory'' system in which they
are derived and conveniently applied to the ``intrinsic'' system, when it
makes sense to define such a system, as is done in the strong coupling
core-particle model.  The resulting theory is completely equivalent to
the starting one and does not yet constitute the standard
phenomenological model.  A second purpose is to describe and implement
the approximation that leads to the standard model.  We describe in
most detail how this may be done in the intrinsic system, the usual 
choice, but emphasize that the approximation may equally be defined
in the laboratory system and that the latter approach has some advantages.

The essential point here may be described as follows.  In the
physical situation, which requires the inclusion of pairing interactions,
the number of solutions of the full KKDF model is twice as great as the 
number of physical states being described.  Hitherto, the  
major technical difficulty (and consumption of cpu time) 
of this method has been the application of a criterion to select the physical 
solutions.  For the ground state problem there is the well-known property
of the BCS theory that the physical solutions (quasiparticles)  
correspond to positive energies and the 
unphysical ones to negative energies.  In the KKDF model the strategy is to
ignore initially rotational excitation energies so as to collapse each
band to a single degenerate state to which the ground state criterion
can be applied.  We then step up the excitation energies, returning
them finally to their
full values; at each step we select the physical solutions by 
a projection technique described in our cited work, that 
involves an extension of the techniques introduced by D\"onau and Frauendorf.

Another way of stating the problem that is directly related to the 
traditional core-particle model is to remark that although only half of the
solutions of the KKDF model are related to physics, the  full set 
of solutions is necessary for mathematical completeness.  
The solutions of our equations at full excitation
can be expanded in terms of the complete set generated at zero excitation,
but this expansion will involve both physical and unphysical states of the
latter limit.
In the conventional core-particle model it is assumed that the 
physical states of the actual problem are well approximated by a superposition
of the physical solutions at zero excitation.  It follows from this that 
it suffices to solve a single eigenvalue problem for the problem of
actual interest rather than having to solve a sequence of such problems.

The third purpose of this paper is to carry through several illustrative
calculations using both the KKDF model and the approximation to it 
just described, in order to assess the  
validity of the latter.  In the course of rethinking
our algorithms in preparation for this study, we have discovered a method 
of simplifying the full calculation to a sufficient extent that much of 
the advantage of technical simplicity of the 
core-particle limit has been wiped out. We shall also describe this new
development.

We start in Secs.\ II and III with a review of the fundamental equations of the
Kerman-Klein method, in order to introduce some improvements in 
notation and presentation, as well as to correct some phase errors 
made previously in the
formulas for transition matrix elements.  In Sec.\ IV we transform our
equations (without approximation) to a description in terms of an intrinsic
frame of reference.  Starting from these equations, the definition and 
formulation of the strong coupling core-particle model 
in its usual form in the
intrinsic system is given in Sec.\ V.   It is explained in Sec.\ VI that an
equivalent and possibly more effective version of this limit can perfectly
well be carried out in the laboratory system.  Turning to applications,
our new algorithm is described in Sec.\ VII and then applied together with the
standard core-particle model to some illustrative cases in Sec.\ VIII.  
Concluding
remarks are presented in Sec.\ IX.  Two appendices provide some technical
details of the derivation carried out in Sec.\ IV.

\section{Fundamental equations of the Kerman-Klein method for
odd nuclei}

We start with a shell-model Hamiltonian of the form
\begin{eqnarray}
H&=& \sum_\alpha h_a a_\alpha^{\dag} a_\alpha  \nonumber \\
&&+ \frac{1}{4} \sum_{abcd}\sum_{LM_L}F_{acdb}(L)
B^{\dag}_{LM_L}(ac)B_{LM_L}(db)  \nonumber \\
&& + \frac{1}{4} \sum_{abcd}\sum_{ML_M} G_{abcd}(L)
A^{\dag}_{LM_L}(ab)A_{LM_L}(cd).     \label{eq:ham}
\end{eqnarray}
Here $h_a$ are the spherical single-particle energies referred to the
nearest closed shell, $\alpha$ refers to the standard set of single-particle
quantum numbers, including in particular the pair $(j_a ,m_a)$ and $a$
refers to the same set with $m_a$ omitted.  $B^{\dag}_{LM_L}$ is the 
particle-hole multipole operator,
\begin{eqnarray}
B^{\dag}_{LM_L}(ab)&\equiv &\sum_{m_a m_b}s_\beta (j_a m_a j_b -m_b|LM_L)
a^{\dag}_\alpha a_\beta  \nonumber \\
&=& (-1)^{j_a +j_b -M_L+1}B_{L-M_L}(ba),
\end{eqnarray}
and $A^{\dag}_{LM_L}$ is the particle-particle multipole operator,
\begin{equation}
A^{\dag}_{LM_L}(ab)\equiv \sum_{m_a m_b} (j_a m_a j_b -m_b|LM_L)
a^{\dag}_\alpha a^{\dag}_{\bar{\beta}},
\end{equation}
where $(j_1 m_1 j_2 m_2|jm)$ is a Clebsch-Gordon (CG) coefficient,
$s_\alpha = (-1)^{j_a -m_a}$, and a bar indicates reversal of the sign  
of the magnetic quantum number.  The coefficients $F$ are the particle-hole
matrix elements,
\begin{equation}
F_{acdb}(L)\equiv \sum_{m's}s_\gamma s_\beta (j_a m_a j_c -m_c|LM_L)
(j_d m_d j_b -m_b|LM_L) V_{\alpha\beta\gamma\delta},
\end{equation}
which satisfies the relation
\begin{equation}
F_{acdb}(L) = (-1)^{j_a +j_b +j_c +j_d} F_{bdca}(L),  \label{eq:F}
\end{equation}
and $G$ the particle-particle matrix elements
\begin{equation}
G_{abcd}(L)\equiv \sum_{m's} (j_a m_a j_b -m_b|LM_L)(j_c m_c j_d -m_d|LM_L)
V_{\alpha\bar{\beta}\gamma\bar{\delta}},
\end{equation}
which satisfies the conditions
\begin{eqnarray}
G_{acdb}(L) &=& (-1)^{j_a +j_c -L +1}G_{cadb} \nonumber \\
&=& (-1)^{j_b +j_d -L +1}G_{acbd}.  \label{eq:G}
\end{eqnarray}

Our initial task is to obtain equations for the states and energies of
an odd nucleus assuming that properties of immediately neighboring even nuclei
are known. The states of the odd nucleus (particle number A) are designated
as $|J \mu\nu\rangle$, where $\nu$ denotes all quantum numbers besides the
angular momentum $J$ and its projection $\mu$. The states of the neighboring
even nuclei with particle numbers $(A\pm 1)$ are written, in a parallel
notation, as $|IMn(A\pm 1\rangle$.  The corresponding eigenvalues are 
$E_{J\nu}$ and $E_{In}^{(A\pm 1)}$, respectively.  The operator equations 
of motion (EOM) are obtained by forming commutators between the single-fermion
operators and the Hamiltonian,
\begin{eqnarray}
{}[a_\alpha,H]&=& h_a^{\prime} a_\alpha   \nonumber \\
&&+ \frac{1}{2} \sum_{bd\gamma}\sum_{LM}s_\gamma (j_a m_a j_c -m_c|LM)
F_{acdb}(L)a_\gamma B_{LM}(db)  \nonumber \\
&&+ \frac{1}{2} \sum_{bd\gamma}\sum_{LM} (j_a m_a j_c -m_c|LM)G_{acbd}(L)
a^{\dag}_{\bar{\gamma}} A_{LM}(bd),  \label{eq:eom1} \\
{}[a^{\dag}_{\bar{\alpha}},H]&=& -h_a^{\prime} a^{\dag}_{\bar{\alpha}} 
\nonumber \\
&&- \frac{1}{2} \sum_{bd\gamma}\sum_{LM}s_{\bar{\gamma}}  
(j_a -m_a j_c m_c|LM)
F_{acdb}(L) B^{\dag}_{LM}(db)a^{\dag}_{\bar{\gamma}}  \nonumber \\
&& -\frac{1}{2} \sum_{bd\gamma}\sum_{LM} (j_a -m_a j_c m_c|LM)
G_{acbd}(L)A^{\dag}_{LM}(bd)a_{\gamma}.  \label{eq:eom2} 
\end{eqnarray}
Here
\begin{equation}
h_a^{\prime} =h_a -\frac{1}{4}\sum_{Lj_c}F_{acac}(L)\frac{2L+1}{2j_a +1}
\end{equation}
are modified single-particle energies.  

The matrix elements of these equations provide expressions that determine the 
single-particle coefficients of fractional parentage,
\begin{eqnarray}
V_{J\mu\nu}(\alpha; IMn)& =& \langle J\mu\nu|a_\alpha|IMn(A+1)\rangle , 
\label{eq:vee} \\
U_{J\mu\nu}(\alpha;IMn)&=& \langle J\mu\nu|a^{\dag}_{\bar{\alpha}}|
IMn(A-1)\rangle.  \label{eq:you}
\end{eqnarray}
To find equations for these quantities, we form the necessary matrix
elements of the EOM and evaluate the interaction terms by inserting the 
completeness relation between the single-fermion operators and the 
multipole or pair operators.  In order to obtain equations that are
expressed completely by means of the amplitudes defined in Eqs.
(\ref{eq:vee}) and (\ref{eq:you}), it is necessary to interchange  
the order of the single-fermion operator and the pair operator in the  
interaction terms of (\ref{eq:eom2}).  This leads to further contributions 
to the single-particle energy in this equation, in that $h_a^{\prime}$ 
is replaced by
$h_a^{\prime\prime}$, with
\begin{equation}
h_a^{\prime\prime} =h_a^{\prime} -\sum_{Lj_c}\frac{2L+1}{2j_a +1}(G_{acac}
+\frac{1}{2}F_{acac}).
\end{equation}

In terms of a convenient and physically
meaningful set of energy differences and sets of multipole fields
and pairing fields defined below, we obtain generalized matrix
equations of the Hartree-Bogoliubov form   
\begin{eqnarray}
{\cal E}_{J\nu}V_{J\mu\nu}(\alpha; IMn) &=&
(\epsilon^{\prime} +\omega^{(A+1)} 
+\Gamma^{(A+1)})_{\alpha IMn,\gamma I^{\prime} M' n'}
V_{J\mu\nu} (\gamma;I' M' n')   \nonumber \\
&&+\Delta_{\alpha IMn,\gamma I' M' n'}U_{J\mu\nu}(\gamma;I'M' n'), 
\label{eq:hfb1} \\
{\cal E}_{J\nu}U_{J\mu\nu}(\alpha; IMn) &=&
(-\epsilon^{\prime\prime} +\omega^{(A-1)} 
-\Gamma^{(A-1)\dag})_{\bar{\alpha} IMn,\bar{\gamma} I' 
M' n'}U_{J\mu\nu}(\gamma;I' M' n')   \nonumber \\
&&-\Delta^{\dag}_{\bar{\alpha} IMn,\bar{\gamma} I' M' n'}
V_{J\mu\nu}(\gamma;I'M' n').   \label{eq:hfb2} 
\end{eqnarray}
Here
\begin{eqnarray}
{\cal E}_{J\nu} &=& -E_{J\nu} +\frac{1}{2}(E_0^{(A+1)}+E_0^{(A-1)}), 
\label{eq:def1}  \\
\epsilon^{\prime}_{\alpha IMn,\gamma I'M'n'} &=& \delta_{\alpha\gamma}
\delta_{II'}\delta_{MM'}\delta_{nn'}(h_a^{\prime} -\lambda_A), 
\label{eq:def2} \\
\lambda_A &=& \frac{1}{2}(E_0^{(A+1)}-E_0^{(A-1)}),  \label{eq:def3} \\
\omega^{(A\pm 1)}_{\alpha INn,\gamma I'M'n'} &=& \delta_{\alpha\gamma}
\delta_{II'}\delta_{MM'}\delta_{nn'}(E_{In}^{(A\pm 1)}-E_0^{(A\pm 1)}),
\label{eq:def4} \\
\Gamma^{(A\pm 1)}_{\alpha IMn,\gamma I'M'n'} &=& \frac{1}{2}\sum_{LM_L}
\sum_{bd}s_\gamma (j_a m_a j_c -m_c|LM_L)F_{acdb}(L), \nonumber \\
&& \langle I'M'n'(A\pm 1)|B_{LM_L}(db)|IMn(A\pm 1)\rangle \label{eq:def5}\\
\Delta_{\alpha IMn,\gamma I'M'n'} &=&\frac{1}{2}\sum_{LM_L}\sum_{bd}
(j_a m_a j_c -m_c|LM_L)G_{acdb}(L) \nonumber \\
&& \langle I'M'n'(A- 1)|A_{LM_L}(db)|IMn(A+1)\rangle. \label{eq:def6}
\end{eqnarray}
Furthermore $E_0^{(A\pm 1)}$ refer to the ground state energies of the 
neighboring even nuclei, the matrix elements of $\Gamma^{\dag}$ 
are derived from those of (\ref{eq:def5})
simply by the replacement of the operator $B$ by $B^{\dag}$, and the matrix
elements of $\Delta^{\dag}$ are similarly derived from those of $\Delta$
by the replacement of $A$ by $A^{\dag}$ together with the interchange
$A\pm 1\rightarrow A\mp 1$.  Finally $\epsilon^{"}_a$ is obtained from
$\epsilon^{'}_a$ by the replacement of $h_a^{'}$ by $h_a^{"}$.

In order to specify a scale for the solutions, we take a suitable matrix element
 of the summed anticommutator,
\begin{eqnarray}
\sum_\alpha \{a_\alpha , a^{\dag}_\alpha\} &=& \Omega, \\
\Omega = \sum_{j_a}(2j_a +1).
\end{eqnarray}
We thus find
\begin{equation}
\frac{1}{\Omega}\sum_{\alpha IMn} [|U_{J\mu\nu}(\alpha; IMn)|^2
+|V_{J\mu\nu}(\alpha; IMn)|^2 =1.  
\end{equation}

All of the above equations are still exact and are not necessarily restricted
to deformed nuclei.  In order to do physics, however, we shall have to impose
restrictions on the number and nature of the core states included in any
application, as well as on the size of the single-particle space.

\section{Matrix elements of single-particle transition operators}

We next apply the formalism to the computation of matrix elements of 
single-particle tensor operators, $T_{LM_L}$, that we write in the form
\begin{equation}
T_{LM_L} =\sum_{\beta\gamma} t_{\beta\gamma}a_\beta^{\dag}a_\gamma.
\label{eq:Tensor}     \end{equation}
The notation is such that the quantities $t_{\alpha\beta}$ include a product
of matrix elements of single-particle operators and of associated coupling
strengths (charges, gyromagnetic ratios, etc.)
We wish to calculate the matrix element $\langle J'\mu'\nu'|T_{LM_L}|
J\mu\nu\rangle$.   To carry through the calculation, we substitute for the 
ket a formally exact expression in terms of the action of single-particle
operators on the states of the core,
\begin{eqnarray}
|J\mu\nu\rangle &=& \frac{1}{\Omega}\sum_{\alpha,IMK}[U_{J\mu\nu}(\alpha,
IMK)a_{\bar{\alpha}}^{\dag}|\underline{IMK}\rangle   \nonumber \\
&& +V_{J\mu\nu}(\alpha,IMK)a_\alpha|\overline{IMK}\rangle],  \label{eq:statev}
\end{eqnarray}
where an underline identifies the lighter of the two cores and an overline 
the heavier one.   
By using the commutation relations and completeness, this leads to the 
following expression for the transition element:
\begin{eqnarray}
\langle J'\mu'\nu'|T_{LM_L}|J\mu\nu\rangle &=&\frac{1}{\Omega}
\sum_{\alpha,IMK,I'M'K'}
[U_{J'\mu'\nu'}(\alpha,I'M'K')U_{J\mu\nu}(\alpha,IMK)  \nonumber \\
&&\times\langle \underline{I'M'K}'|T_{LM_L}|\underline{IMK}\rangle    
\nonumber \\
&+& [V_{J'\mu'\nu'}(\alpha,I'M'K')V_{J\mu\nu}(\alpha,IMK)
\langle \overline{I'M'K'}|T_{LM_L}|\overline{IMK}\rangle    \nonumber \\
&+& \frac{1}{\Omega}\sum_{\alpha,\gamma,IMK}t_{\alpha\gamma}[U_{J'\mu'\nu'}
(\bar{\alpha},IMK)U_{J\mu\nu}(\bar{\gamma},IMK)  \nonumber \\
&& - V_{J\mu\nu}(\alpha,IMK)V_{J'\mu'\nu'}(\gamma,IMK)].  \label{eq:tran}
\end{eqnarray}

This is now evaluated by use of the Wigner-Eckart theorem with the following
definitions of the reduced matrix elements:
\begin{eqnarray}
\langle J'\mu'\nu'|T_{LM_L}|J\mu\nu\rangle &=&\frac{(-1)^{J-\mu}}
{\sqrt{2L+1}}(J'\mu'J-\mu|LM_L)\langle J'\nu'||T_L||J\nu\rangle,\label{eq:T} \\
\langle I'M'K'|T_{LM_L}|IMK\rangle &=& \frac{(-1)^{I-M}}{\sqrt{2L+1}}
(I'M'I-M|LM_L)    \nonumber \\
&& \times\langle I'K'||T_L||IK\rangle,  \label{T1} \\
t_{\alpha\gamma} &=& \frac{(-1)^{j_c -m_c}}{\sqrt{2L+1}}
(j_a m_a j_c -m_c|LM_L) t_{ac},        \label{T2}
\end{eqnarray}
\begin{eqnarray}
V_{J\mu\nu}(\alpha,IMK)&=&\frac{(-1)^{J-\mu}}{\sqrt{2j_a +1}}
(IMJ-\mu|j_a m_a)v_{J\nu}(aIK),  \label{eq:WE1} \\
U_{J\mu\nu}(\alpha,IMK)&=&\frac{(-1)^{J-\mu +j_a +m_a}}{\sqrt{2j_a +1}}
(IMJ-\mu|j_a m_a)u_{J\nu}(aIK).  \label{eq:WE2} 
\end{eqnarray}
With the help of these definitions, we obtain the formula for the 
reduced matrix element that is utilized in the KKDF model.
\begin{eqnarray}
\langle J'\nu'||T_L||J\nu\rangle &=& \frac{1}{\Omega}\sum_{aIKI'K'}
(-1)^{j_a +J'+I+L}\left\{\begin{array}{ccc}I & I' & L \\ J' & J & j_a
\end{array}\right\}   \nonumber \\
&& \times [u_{J\nu}(aIK)u_{J'\nu'}(aI'K')\langle \underline{I'K'}||T_L||
\underline{IK}\rangle  \nonumber \\
&& + v_{J\nu}(aIK)v_{J'\nu'}(aI'K')\langle \overline{I'K'}||T_L||
\overline{IK}\rangle]  \nonumber \\
&+& \frac{1}{\Omega}\sum_{acIK}t_{ac}
[(-1)^{j_a +I+J+L}\left\{\begin{array}{ccc}j_a &j_c&L\\ J& J' &I
\end{array}\right\}u_{J'\nu'}(aIK)u_{J\nu}(cIK)  \nonumber \\
&&+(-1)^{j_a +I+J+1}\left\{\begin{array}{ccc}j_a &j_c&L\\ J'& J &I
\end{array}\right\}v_{J\nu}(aIK)v_{J'\nu'}(cIK)].    \label{eq:RME}
\end{eqnarray}
This is, with some phase corrections, the formula 
that was derived in a previous work.

\section{Transformation to intrinsic system for axial case}

We have described previously \cite{pav:1,pav:2,pav:3} 
several applications of the formalism
reviewed in the preceding sections to strongly deformed nuclei. Some of the
results, together with some additional calculations, will be used as the 
basis for a numerical study of the relation of the method of this paper
to the traditional strong coupling core-particle model. As will
be explained in Sec.\ VI, this relation can be studied using
the formalism already at hand (theory expressed in the ``laboratory''
system of coordinates); in fact it turned out to be  
economical for us to carry out
all numerical work from this standpoint. Nevertheless, in the following
two sections we shall undertake to develop  the connection between our
method and the way such calculations are normally presented in the 
intrinsic system.  Our justification for this digression is that whenever
we have presented a public account of our previous work in this field,
one question invariably raised was precisely this connection.  In what
follows, we shall answer the question raised in two steps.  In the first, we
shall derive a form of our equations in the intrinsic coordinate
system that is fully equivalent to the theory described above.  Second
we shall show that the conventional core-particle approach involves
a further specialization of this general result and examine this limiting
case in some theoretical detail.

For illustrative purposes, we
take a model of the even (core) nuclei that consists of the ground-state
band $|IMK=0\rangle=|IM\rangle$ and a finite number of positive parity excited
bands $|IMKn\rangle$.  For the remainder of this section the symbol $n$
will be suppressed.  We are thus assuming that the eigenstates of the
even nuclei have axial symmetry and that their eigenstates can be assigned
a definite value of $K$, the component of the angular momentum along the 
figure axis.  This assumption is reasonable as long as the 
states of the same angular momentum belonging to different bands  
are well-separated in energy.

We first use rotational invariance to study the structure of the 
amplitudes $V$ and $U$ defined in Eqs.\ (\ref{eq:vee}) and (\ref{eq:you}),
respectively.  For this purpose we introduce a complete set of states
$|R\rangle$ localized in the Euler angles, $R=(\alpha\beta\gamma)$
and write
\begin{eqnarray}
|IMK\rangle &=& \int dR\, |R\rangle\langle R|IMK\rangle \nonumber \\
&=& \left(\frac{2I+1}{8\pi^2}\right)^\frac{1}{2} \int dR \,
|R\rangle D^{(I)}_{MK}(R) 
\label{eq:Wigner}    \end{eqnarray}
The identification
of a scalar product of many-body states with the Wigner $D$ function
is part of the definition of the model.
When (\ref{eq:Wigner}) is substituted into the definition of $V$, 
and use is made of the definitions to follow, we are 
thereby led to the study of an amplitude such as
\begin{eqnarray}
\langle J\mu\nu|a_\alpha |R\rangle &=& \langle J\mu\nu|U(R)U^{-1}(R)
a_\alpha U(R)|0\rangle \nonumber \\
&=&\sum_{\mu'\kappa_a} \langle J\mu\nu|U|J\mu'\nu\rangle
\langle J\mu'\nu|U^{-1}a_\alpha U|0\rangle   \nonumber \\
&=&\sum_{\mu'\kappa_a} D^{(J)\ast}_{\mu\mu'}(R)D^{(j_a)\ast}_{m_a\kappa_a }(R)
\chi_{J\mu'\nu} (j_a\kappa_a)(-1)^{j_a +\kappa_a},    \label{eq:chi1}
\end{eqnarray}
where $U(R)$ is a unitary rotation operator defined by the value of $R$.
The previous manipulations have utilized the following relations 
and definitions (of which the first two are standard):
\begin{eqnarray}
\langle JK|U(R)|JM\rangle &=& D^{(J)\ast}_{KM}(R), \label{eq:wigner} \\
U^{-1}(R)a_{jm}U(R) &=& \sum_{\kappa}a_{j\kappa}D^{(j)\ast}_{m\kappa}(R), 
\label{eq:tensor} \\
\langle J\mu\nu|a_{jm}|0\rangle&\equiv&(-1)^{j+m}\chi_{J\mu\nu}(jm)
\label{eq:chi2}
 \end{eqnarray}
The introduction of the phase in (\ref{eq:chi2}) simplifies the structure
of the transformed equations of motion given below.

With the help of the integral of a product of three $D$ functions
and the application of standard symmetry properties of CG coefficients,
we find 
\begin{eqnarray}
V_{J\mu\nu} (\alpha; IMK) &=& \sum_{\kappa_a}\sqrt{\frac{8\pi^2}{2j_a +1}}
(-1)^{J-\mu}  (IMJ-\mu|j_a m_a)     \nonumber \\
&& \times (JK-\kappa_a j_a\kappa_a|IK)(-1)^{j_a +\kappa_a}
\chi_{JK-\kappa_a\nu} (j_a\kappa_a) . \label{eq:xvee}
\end{eqnarray}
A similar analysis carried out for the amplitude $U$ yields the result
\begin{eqnarray}
U_{J\mu\nu}(\alpha; IMK) &=&\sum_{\kappa_a} \sqrt{\frac{8\pi^2}{2j_a +1}}
(-1)^{J-\mu +j_a -\kappa_a +j_a +m_a}(IMJ-\mu|j_a m_a) \nonumber \\
&&\times(JK-\kappa_a j_a \kappa_a|IK) 
\phi_{JK-\kappa_a\nu}(j_a\kappa_a),   \label{eq:xyou}  \\ 
\phi_{J\mu\nu} (j_a\kappa_a) &=& \langle J\mu\nu|a^{\dag}_{j_a -\kappa_a}
|0\rangle.   \label{eq:deffi} 
\end{eqnarray}

Starting from Eqs. (\ref{eq:hfb1}) and (\ref{eq:hfb2}) and utilizing the
forms (\ref{eq:xvee}) and (\ref{eq:xyou}), we next derive equations
satisfied by the amplitudes $\chi$ and $\phi$.  The technique is to 
eliminate the CG coefficients that occur in (\ref{eq:xvee}) and 
(\ref{eq:xyou}) by multiplying by $(IMJ-\mu|j_a m_a)
(JK-\kappa_a j_a\kappa_a|IK)$
and by the reciprocal of the factors pre-multiplying these CG coefficients
in the one or the other of these equations, 
summing over $M,\mu$ and $I$, and using standard
formulas of angular momentum algebra.  Some details are provided in Appendix
A.  In the equations to follow, the quantities that appear for the first 
time are defined by the equations
\begin{eqnarray}
{\cal R}(m,K|j,J) &=& \sqrt{(j+m)(j-m+1)} \nonumber \\
&&\times\sqrt{(J-K+m)(J+K-m+1)}, \label{eq:CC} \\
\langle I'M'K'|B_{LM_L}^{\dag}(db)|IMK\rangle &=& 
q_{K'K}^{(L,0)}(db)\sqrt{\frac{2I+1}{2I' +1}} \nonumber \\
& \times&  (IMLM_L|I'M')(IKLK'-K|I'K'), 
\label{eq:q} \\
\langle I'M'K'|A_{LM_L}^{\dag}(db)|IMK\rangle &=& 
\Delta_{K'K}^{(L,0)}(db)\sqrt{\frac{2I+1}{2I' +1}} \nonumber \\
& \times&  (IMLM_L|I'M') (IKLK'-K|I'K'), 
\label{eq:d} \\
\omega_{IK}^{(A\pm 1)}& =&E_K^{(A\pm 1)} +\frac{1}{2{\cal I}_K^{(A\pm 1)}}
[I(I+1) -K^2].    \label{eq:X}
\end{eqnarray}
Of these equations, the quantity ${\cal R}$ is recognized as arising from
the matrix elements of the Coriolis coupling and the remaining equations
are expressions valid for the axial rotor model for matrix elements
of transition operators (see further below) and excitation energies.  
These expressions constitute
definitions of the intrinsic multipole  moments $q$, of the intrinsic
pairing moments $\Delta$, of the band-head energies $E_K$, and of the 
moments of inertia ${\cal I}_K$.

The resulting equations (with partial suppression of the index $\nu$) are
\begin{eqnarray}
{\cal E}_{J\nu}\chi_{J,K-\kappa_a}(j_a\kappa_a)& =&
\{\epsilon_a^{\prime} +E_K^{(A+1)}+\frac{1}{2{\cal I}_K^{(A+1)}}
[J(J+1) -K^2  \nonumber\\
&&+j_a (j_a+1) +2\kappa_a(K-\kappa_a)]\} 
\chi_{J,K-\kappa_a}(j_a\kappa_a)   \nonumber \\
&&+\frac{1}{2{\cal I}_K^{(A+1)}}{\cal R}(\kappa_a,K|j_a,J)
\chi_{J,K-\kappa_a +1}(j_a\kappa_a -1) \nonumber \\
&&+\frac{1}{2{\cal I}_K^{(A+1)}}{\cal R}(-\kappa_a,-K|j_a,J)
\chi_{J,K-\kappa_a -1}(j_a\kappa_a +1) \nonumber \\
&&+\sum_{bcd\kappa_c K'L}\frac{1}{2}(-1)^{j_c +\kappa_a +L} 
 F_{acdb}(L)q^{(L,0)}_{K'K}(db)\nonumber \\
&&\times(j_c -\kappa_c j_a\kappa_a|LK-K')\chi_{J,K-\kappa_a}(j_c\kappa_c)
\nonumber \\
&&+\sum_{bcd\kappa_c K'L}\frac{1}{2}                        
(-1)^{j_c  +\kappa_a +L}G_{acdb}(L) 
\Delta^{(L,0)}_{K'K}(db) \nonumber \\
&&\times(j_c -\kappa_c j_a \kappa_a|LK-K')\phi_{J,K-\kappa_a}(j_c\kappa_c), 
\label{eq:cp1}  \\
{\cal E}_{J\nu}\phi_{J,K-\kappa_a}(j_a\kappa_a)& =&
\{-\epsilon_a^{\prime\prime} +E_K^{(A-1)}+\frac{1}{2{\cal I}_K^{(A-1)}}
[J(J+1) -K^2   \nonumber\\
&&+j_a (j_a+1) +2\kappa_a(K-\kappa_a)]\} 
\phi_{J,K-\kappa_a}(j_a\kappa_a)   \nonumber \\
&&+\frac{1}{2{\cal I}_K^{(A-1)}}{\cal R}(\kappa_a,K|j_a,J) 
\phi_{J,K-\kappa_a +1}(j_a\kappa_a -1) \nonumber \\
&&+\frac{1}{2{\cal I}_K^{(A-1)}}{\cal R}(-\kappa_a,-K|j_a,J)
\phi_{J,K-\kappa_a -1}(j_a\kappa_a +1) \nonumber \\
&&-\sum_{bcd\kappa_c K'L}\frac{1}{2}
(-1)^{j_c +\kappa_c} F_{acdb}(L) 
q^{(L,0)}_{K'K}(db) \nonumber \\
&&\times(j_c -\kappa_c j_a\kappa_a|LK-K')
\phi_{J,K-\kappa_a}(j_c\kappa_c)  \nonumber \\
 &&+\sum_{bcd\kappa_c K'L}\frac{1}{2}
G_{acdb}(L) \Delta^{(L,0)}_{K'K}(db) 
(-)^{j_c +\kappa_c}  \nonumber \\
&&\times(j_c -\kappa_c j_a\kappa_a|LK-K')
\chi_{J,K-\kappa_a}(j_c\kappa_c).  \label{eq:cp2}
\end{eqnarray}
In these expressions, we have deliberately chosen, for conciseness of  
expression, not to do the sum on $\kappa_c$, where the value
$\kappa_c = K'-K-\kappa_a$ is imposed by the resident CG coefficient.

Relations (\ref{eq:q}) and (\ref{eq:d}), which have been used in all our 
previous applications, are approximate, and therefore
require further discussion.  
For example, Eq.\ (\ref{eq:q}) follows as the value of the first term of 
the operator expression
\begin{eqnarray}
B_{LM_L}^{\dag}(db)& =& \sum_{\lambda_1,...,\lambda_p}
\sum P_{I'M'K'}q^{(L,\lambda)}_{K'K}(db)   \nonumber \\
&&\times \{D^{(L)}_{M_L,K'-K+\lambda_1 +...+\lambda_p},I'_{\lambda_1}...
I'_{\lambda_p}\}P_{IMK}.   \label{eq:b}
\end{eqnarray}
Here the $P$ are the projection operators for the specified band
members, $I'_\lambda$
is a spherical tensor component of the intrinsic angular momentum, and the 
braces imply a 
symmetrized expression.   Assuming that the connected bands have the same
parity, p is even for even electric multipoles and odd magnetic multipoles
and odd for odd electric multipoles and even magnetic multipoles. If the
connected bands have opposite parity, there is a corresponding relation.
The form of (\ref{eq:b})
is a consequence of the assumption that $B$ must be a tensor operator
of appropriate rank in the Hilbert space of the axial rotor.  The 
further assumption
that we can limit ourselves to the first term is that for the states of 
interest the rotor is almost rigid, as is true for the low-lying states
of strongly deformed nuclei.  The corresponding expression for the pairing
operator requires only the replacements
\begin{eqnarray}
q^{(L,\lambda)}_{K'K} &\rightarrow& \Delta^{(L,\lambda)}_{K'K},   
\end{eqnarray}
and the realization that the projection operators to the left and to the 
right refer to different cores.  

The inclusion of odd multipole or pairing interactions requires that, 
minimally, we
choose $p=1$.  The evaluation of such a multipole term is carried
out in Appendix B.

\section{Core-particle coupling model}
\subsection{Spectra}

For further development, we specialize the formulas of the previous section 
to the conventional
monopole pairing plus quadrupole-quadrupole model and confine our attention  
initially to the special case that we include only the ground-state
band of the neighboring even cores.  (The general case will be considered
subsequently.)  We also assume that we are treating well-deformed nuclei  
and ignore number conservation.   For $L=0$ pairing we have in the limit 
of a constant pairing matrix element
\begin{eqnarray}
-\sum_b G_{aabb}\Delta^{(0,0)}_{00}(bb) &\equiv& 2\Delta_a\sqrt{2j_a +1}
\nonumber \\
 &\cong & 2\Delta\sqrt{2j_a +1}.  \label{eq:delta}
 \end{eqnarray}
For the quadrupole interaction, we write
\begin{eqnarray}
F_{abcd}(2) &=& -\kappa_2 F_{ab}F_{dc}, \label{eq:F1} \\
\sum_{bd}F_{db}q^{(2,0)}_{00}(db) & \equiv & Q_0. \label{eq:F2}
\end{eqnarray}
Because we are dealing with a $K=0$ band, axial symmetry implies that
$\kappa_a =\kappa_c =\kappa$, and the quadrupole potential becomes
\begin{equation}
{\cal V}_{ac}^\kappa = -\frac{1}{2}\kappa_2 F_{ac}Q_0 (-1)^{j_c +\kappa}
(j_c -\kappa j_a\kappa|20).   \label{eq:Vee}
\end{equation}
The potential ${\cal V}$ is symmetric provided we choose 
\begin{equation}
F_{ca} = (-1)^{j_a +j_c +1}F_{ac}, 
\end{equation}
which is consistent with (\ref{eq:F}).

We next study the limit of our equations found by introducing the 
simplifications made above and also neglecting the core excitation energies.
The resulting equations do not depend on the total angular momentum,
and we thus set (with $\kappa_a =\kappa$) 
\begin{eqnarray}
\chi_{J,-\kappa}(j_c\kappa)&\rightarrow& \chi_{\kappa c}, 
\nonumber \\
\phi_{J,-\kappa}(j_c\kappa)&\rightarrow& \phi_{\kappa c}, 
\nonumber \\ 
{\cal E}_{J\nu} &\rightarrow & {\cal E}_{\kappa\tau}.  \label{eq:arrow}
\end{eqnarray}
Evidently $\kappa$ is the component of the quasi-particle angular momentum
along the axis of symmetry, and $\tau$ resolves degeneracies in the values of
$\kappa$.
In the limit considered our equations thus reduce to a Hartree-Bogoliubov set
\begin{eqnarray}
{\cal E}_{\kappa\tau}\chi_{\kappa a}&=&\epsilon_a\chi_{\kappa a}
+{\cal V}_{ac}^\kappa\chi_{\kappa a} -\Delta\phi_{\kappa a}, \label{eq:HB1}\\
{\cal E}_{\kappa\tau}\phi_{\kappa a}&=&-\epsilon_a\phi_{\kappa a}
-{\cal V}_{ac}^\kappa\phi_{\kappa a} -\Delta\chi_{\kappa a}. \label{eq:HB2} 
\end{eqnarray}
>From now on we set $\epsilon_a^{\prime} =\epsilon_a^{\prime\prime} =\epsilon_a$.

These equations are solved by introducing the unitary transformation that
diagonalizes the single-particle Hamiltonian
\begin{equation}
{\cal H}_{ac}^\kappa = \epsilon_a\delta_{ac}+{\cal V}_{ac}^\kappa,  
\label{eq:sph}
\end{equation}
namely,
\begin{eqnarray}
\chi_{\kappa c} &=& \sum_{\tau} A_{c\tau}^\kappa v_{\kappa\tau}, \nonumber \\
v_{\kappa\tau} &=& \sum_c A_{c\tau}^{\kappa\ast}\chi_{\kappa c}, 
\label{eq:trans} \\
\sum_\tau A_{a\tau}^{\kappa\ast}A_{b\tau}^\kappa 
& =& \delta_{ab}, \nonumber \\
\sum_a A_{a\tau}^{\kappa\ast}A_{a\tau'}^\kappa 
&=& \delta_{\tau\tau'}, \nonumber \\
\sum_{ac}A_{a\tau}^{\kappa\ast}{\cal H}_{ac}^\kappa A_{c\tau'}^\kappa 
&=& e_{\kappa\tau}\delta{\tau\tau'}.
\end{eqnarray}
We thus obtain a standard set of BCS equations 
\begin{eqnarray}
{\cal E}_{\kappa\tau}v_{\kappa\tau} &=& e_{\kappa\tau}v_{\kappa\tau}                        
-\Delta u_{\kappa\tau}, \label{eq:BCS1} \\
{\cal E}_{\kappa\tau}u_{\kappa\tau} &=&- e_{\kappa\tau}u_{\kappa\tau}               
-\Delta v_{\kappa\tau}, \label{eq:BCS2}  
\end{eqnarray}
with the usual solutions
\begin{equation}
{\cal E}_{\kappa\tau} = \pm \sqrt{e_{\kappa\tau}^2 + \Delta^2},
\end{equation}
where corresponding to the plus sign, we have the physical solutions
\begin{equation}
\psi_{\kappa\tau} =\left( \begin{array}{c} v_{\kappa\tau} \\  u_{\kappa\tau}
 \end{array}\right) ,
\end{equation}
and to the minus sign the unphysical solutions
\begin{equation}
\bar{\psi}_{\kappa\tau}=\left(\begin{array}{c}-u_{\kappa\tau}\\v_{\kappa\tau}
\end{array}\right).
\end{equation}
We have reviewed this familiar material because of its importance in the 
definition of the standard core-particle model.

We have now laid the groundwork for the solution of the full equations of
motion (\ref{eq:cp1}) and (\ref{eq:cp2}).  For this general solution 
the notational change contained in (\ref{eq:arrow}) is  generalized to
\begin{eqnarray}
\chi_{J,-\kappa}(j_c\kappa)&\rightarrow& \chi_{J\kappa c}, 
\nonumber \\
\phi_{J,-\kappa}(j_c\kappa)&\rightarrow& \phi_{J\kappa c}, 
\nonumber \\ 
{\cal E}_{J\nu} &\rightarrow & {\cal E}_{J\kappa\tau}.  \label{eq:arrow1}
\end{eqnarray}
Introducing again the transformation that 
diagonalizes the single-particle Hamiltonian ${\cal H}^\kappa$, 
\begin{eqnarray}
\chi_{J\kappa a} &=& \sum_\tau A_{a\tau}^\kappa
\chi_{J\kappa\tau}, \nonumber \\
\chi_{J\kappa\tau} &=& \sum_a A_{a\tau}^{\kappa\ast}
\chi_{J\kappa a},
\end{eqnarray}
with a corresponding transformation for $\phi$, the equations of motion become
\begin{eqnarray}
{\cal E}_{J\kappa\tau}\chi_{J\kappa\tau} &=&e_{\kappa\tau}
\chi_{J\kappa\tau} -\Delta\phi_{J\kappa\tau} \nonumber\\
&&+ \sum_{\kappa'\tau'}U_{\kappa\tau,\kappa'\tau'}^J 
\chi_{J\kappa'\tau'},    \label{eq:cp3}  \\
{\cal E}_{J\kappa\tau}\phi_{J\kappa\tau} &=&-e_{\kappa\tau} 
\phi_{J\kappa\tau} -\Delta\chi_{J\kappa\tau} \nonumber\\
&&+ \sum_{\kappa'\tau'}U_{\kappa\tau,\kappa'\tau'}^J 
\phi_{J\kappa'\tau'},      \label{eq:cp4}
\end{eqnarray}
and the non-vanishing matrix elements of $U$ that occur in these 
equations (that reinstate the angular momentum and include the 
Coriolis coupling) are
\begin{eqnarray}
2{\cal I}U_{\kappa\tau,\kappa\tau'}^J &=& \sum_a A_{a\tau}^{\ast}
[J(J+1) +j_a(j_a +1)- 2\kappa^2 ]A_{a\tau'}, \nonumber \\
2{\cal I}U_{\kappa\tau,\kappa -1 \tau'}^J &=& \sum_a A_{a\tau}^{\ast}
{\cal R}(\kappa,0|j_a,J)A_{a\tau'},  \nonumber \\
2{\cal I}U_{\kappa\tau,\kappa +1 \tau'}^J &=& \sum_a A_{a\tau}^{\ast}
{\cal R}(-\kappa,0|j_a,J)A_{a\tau'}.  \label{eq:U}
\end{eqnarray}
We recall that the quantities ${\cal R}$ are defined in (\ref{eq:CC}).

At this point we introduce the defining  approximation for the core-particle
coupling model, first defining
\begin{equation}
\Psi_{J\kappa\tau}=\left(\begin{array}{c}\chi_{J\kappa\tau}
 \\ \phi_{J\kappa\tau} \end{array}\right),
\end{equation}
and then assuming that $\Psi$ can be expanded in terms of the physical 
solutions of Eqs.\ (\ref{eq:HB1}) and (\ref{eq:HB2}), namely,
\begin{equation}
\Psi_{J\kappa\tau}\cong C_{J\kappa\tau}\psi_{\kappa\tau}. 
\label{eq:approx:}
\end{equation}
By contrast, the exact expression must be the 
sum of a physical and an unphysical
solution.  If we include the latter, we have, in fact, returned to the 
KKDF model and to its basic technical problem of selecting physical 
solutions. Although, as we shall see later, we have found a simplified method
to handle this problem, compared to the approach used in earlier work,
it remains of interest to know when the traditional core-particle model is 
valid.

With the help of (\ref{eq:approx:}), Eqs.\ (\ref{eq:cp3}) and (\ref{eq:cp4})
can be reduced to a form of the core-particle coupling equations 
ready for final numerical study, namely,
\begin{eqnarray}
{\cal E}_{J\kappa\tau}C_{J\kappa\tau} & =& {\cal E}_{\kappa\tau}
C_{J\kappa\tau} +\sum_{\kappa'\tau'}W_{\kappa\tau,\kappa'\tau'}^J 
C_{J\kappa'\tau'},  \label{eq:final} \\
W_{\kappa\tau,\kappa'\tau'} &=& \tilde{\psi}_{\kappa\tau}
U_{\kappa\tau,\kappa'\tau'}^J \psi_{\kappa'\tau'}.
\end{eqnarray}
This is a standard diagonalization problem with the ``correct" number of 
solutions.  These solutions will be compared with the exact solution of the
corresponding KKDF equations.

We consider next the general case defined in the theoretical
formulation of the previous section, with multiple bands in the core nuclei, 
but with the maintenance
of axial symmetry.  Though not really necessary, it makes sound physical
sense to proceed as follows: We lean on the fact that the interband quadrupole
transitions are weak compared to intraband transitions.  Thus we shall 
first ignore the terms associated with these transitions as well as the 
perturbation associated with finite excitation energy above the band-head.  
What remains is a Hartree-Bogoliubov approximation for excited bands.
Next we add the ``Coriolis coupling" and thus obtain a series of bands
in close analogy with our treatment of Coriolis coupling for the 
ground-state band.  Finally, we introduce the coupling arising from
interband transitions in the cores.  

In fact, it is hardly necessary to give many details of the previous steps.
All we need is an enhanced notation.  Instead of the ground state band, we 
consider a band $K\sigma$, where $00$ is the ground state band, $01$  the 
beta band, $20$ the gamma band, etc.  Now to all the quantities defined 
above, such as $\Delta$, $Q_0$, ${\cal V}_{ac}^\kappa$, $\chi_{\kappa c}$,
$\phi_{\kappa c}$, etc., we add a superscript $(K\sigma)$.  Thus after 
transformation by the matrix $A_{c\tau}^{\kappa,K\sigma}$ the excited 
state HB equations become
\begin{eqnarray}
({\cal E}_{\kappa\tau}^{K\sigma}-E^{K\sigma})v_{\kappa\tau}^{K\sigma} &=& 
e_{\kappa\tau}^{K\sigma}v_{\kappa\tau}^{K\sigma}                        
-\Delta^{K\sigma} u_{\kappa\tau}^{K\sigma}, \label{eq:BCS3} \\
({\cal E}_{\kappa\tau}^{K\sigma}-E^{K\sigma})u_{\kappa\tau}^{K\sigma} &=& 
-e_{\kappa\tau}^{K\sigma}u_{\kappa\tau}^{K\sigma}                        
-\Delta^{K\sigma} v_{\kappa\tau}^{K\sigma}, \label{eq:BCS4} 
\end{eqnarray}
with the solutions
\begin{equation}
{\cal E}_{\kappa\tau}^{K\sigma} =E^{K\sigma} \pm \sqrt{(e_{\kappa\tau}
^{K\sigma})^2 + (\Delta^{K\sigma})^2},
\end{equation}
where the first term on the right hand side is clearly the band-head energy.
The remainder of the calculation also parallels that made for the case of the 
ground-state band.  The only quantities requiring more than a notational 
change are the matrix elements of the operator $U$ defined in Eq.\
(\ref{eq:U}).  The necessary emendations can be read off directly form
the core-particle equations (\ref{eq:cp1}) and (\ref{eq:cp2}).

Thus we have specified a procedure for deriving a set of state vectors
$\Psi_{J\nu}^{K\sigma}$ and associated energies ${\cal E}_{J\nu}^{K\sigma}$,
where we have amalgamated the pair of quantum numbers ${\kappa\tau}$
into the symbol $\nu$.  We have taken account of all terms in the effective
Hamiltonian except for the interband multipole fields.  To finally
include the latter, we write 
\begin{equation}
\hat{{\cal V}}=\tau_3(\hat{{\cal V}}_d +\hat{{\cal V}}_{od}), 
\end{equation}
where $\tau_3$ is the usual Pauli matrix, $d$ refers to the intraband parts
of the multipole field and $od$ to the interband parts.  It remains 
to take into account only the latter piece.  This is done by a final
mixing
\begin{equation}
\Theta_{J\rho} = \sum_{\nu K\sigma}{\cal A}^J_{\rho,\nu K\sigma}
\Psi_{J\nu}^{K\sigma},    
\end{equation}
where the mixing coefficients are determined by the conditions
\begin{eqnarray}
{\cal E}_{J\rho}{\cal A}^J_{\rho,\nu K\sigma}&=& {\cal E}^{K\sigma}_{J\nu}
{\cal A}_{\rho,\nu K\sigma}^J  \nonumber \\
&&+ \sum_{\nu' K'\sigma'}{\cal F}^J_{\nu K\sigma,\nu' K'\sigma'}
{\cal A}^J_{\rho, \nu' K'\sigma'}, \\
{\cal F}^J_{\nu K\sigma,\nu' K'\sigma'}&=& \tilde{\Psi}^{K\sigma}_{J\nu}
\tau_3\hat{{\cal V}}_{od}\Psi^{K'\sigma'}_{J\nu'}.
\end{eqnarray}

In the last two sections, we have derived the conventional form of the 
core-particle coupling theory from the KKDF formalism.  In fact the equations
derived in the first of these sections were exact, 
i.\ e., completely equivalent to those
of KKDF, indeed only their form in the ``intrinsic frame".  The core-particle
coupling model as customarily presented involves, as described, an
additional approximation in the solution of these equations.  Indeed, the 
essence of the model lies in this approximation rather than in whether
calculations are carried out in the intrinsic system as described above
or in the laboratory system as is done in the full application of the 
KKDF method.   In the next full section, we shall record the form 
of the core-particle approximation in the intrinsic system.
coupling theory in the laboratory frame.

\subsection{Core-particle coupling model: transitions}

Here we shall only indicate the step involving the exact transformation
of (\ref{eq:RME})  into an expression referring to the intrinsic system.
Since we shall not utilize this version of the formalism, we leave the 
further transformation by the introduction of the approximate solutions
developed in the preceding subsection as an exercise for the reader.
This step is to introduce values for the reduced matrix elements on the 
right hand side of (\ref{eq:RME}) and to carry out the summations over 
$I$ and $I'$ in order to reach a formula appropriate to the core-particle
coupling model.  By comparing (\ref{eq:T}) with (\ref{eq:q}), (\ref{eq:WE1})
with (\ref{eq:xvee}), and (\ref{eq:WE2}) with (\ref{eq:xyou}), we can
read off the formulas
\begin{eqnarray}
\langle I'K'||T_L||IK\rangle &=& \sqrt{2I+1}q_{K'K}^{(L,0)}
(IKLK'-K|I'K'),       \label{eq:RMT} \\
v_{J\nu}(aIK)&=&\sum_{\kappa_a}\sqrt{8\pi^2}(-1)^{j_a +\kappa_a}
(JK-\kappa_a j_a \kappa_a|IK) \nonumber \\
&&\times \chi_{JK-\kappa_a}(j_a\kappa_a),  \label{eq:RMV} \\
u_{J\nu}(aIK)&=&\sum_{\kappa_a}\sqrt{8\pi^2}(-1)^{j_a -\kappa_a}
(JK-\kappa_a j_a \kappa_a|IK)  \nonumber \\
&& \times \phi_{JK-\kappa_a}(j_a\kappa_a). \label{eq:RMU} 
\end{eqnarray}
Carrying out the summations over $I$ and $I'$, we are led to the equation
\begin{eqnarray}
\langle J'\nu'||T_L||J\nu\rangle = \frac{8\pi^2}{\Omega}\sum_{a\kappa_a KK'}
\frac{1}{\sqrt{2J' +1}}(JK-\kappa_a LK'-K|J'K'-\kappa_a) &&\nonumber \\
\times [\phi_{JK-\kappa_a}(j_a\kappa_a)\phi_{J'K'-\kappa_a}(j_a \kappa_a)
+\chi_{JK-\kappa_a}(j_a\kappa_a)\chi_{J'K'-\kappa_a}(j_a \kappa_a)] &&
\nonumber \\                                                      
+\frac{8\pi^2}{\Omega}\sum_{acK}t_{ac} 
 [(-1)^{j_c +\kappa_c}(j_a -\kappa_a j_c\kappa_c|L\kappa_c -\kappa_a)
(JK-\kappa_c L\kappa_c -\kappa_a|J'K-\kappa_a)&& \nonumber \\
\times \frac{1}{\sqrt{(2L+1)(2J'+1)}}\phi_{J'K-\kappa_a}(j_a\kappa_a)
\phi_{JK-\kappa_c}(j_c\kappa_c) &&   \nonumber \\
+(-1)^{j_c +\kappa_c +J+J'+L}(j_a -\kappa_a j_c\kappa_c|L\kappa_c -\kappa_a)
(J'K-\kappa_c L\kappa_c -\kappa_a|JK-\kappa_a)&& \nonumber \\
\times \frac{1}{\sqrt{(2L+1)(2J+1)}}\chi_{JK-\kappa_a}(j_a\kappa_a)
\chi_{J'K-\kappa_c}(j_c\kappa_c).&&   \label{eq:TCPCM} 
\end{eqnarray}

\section{Core-particle coupling model in laboratory frame}
\label{cplab}
We show here that the core-particle coupling model can be formulated just as
conveniently in the laboratory frame of reference as in the intrinsic
frame.  We start with the fundamental matrix equations of motion,
Eqs.\ (\ref{eq:hfb1}) and (\ref{eq:hfb2}) and reduce them by application
of the Wigner-Eckart theorem.  
By means of Eqs.\ (\ref{eq:WE1}) and (\ref{eq:WE2}), 
Eqs.\ (\ref{eq:q}) and (\ref{eq:d}), and 
standard angular momentum algebra, we find the equations (assuming that
$K' -K$ and $L$ are even, as is the case for the specific model considered
in the body of this paper),
\begin{eqnarray}
{\cal E}_{J\nu}v_{J\nu}(aIK) &=& (\epsilon_a +\omega_{IK})v_{J\nu}(aIK)
\nonumber \\
&& +\sum_{cI'K'} \Gamma(aIK,cI'K')v_{J\nu}(cI'K') \nonumber \\
&& +\sum_{cI'K'} \Delta(aIK,cI'K')u_{J\nu}(cI'K'), \label{eq:red1} \\
{\cal E}_{J\nu}u_{J\nu}(aIK) &=& (-\epsilon_a +\omega_{IK})u_{J\nu}(aIK)
\nonumber \\
&& -\sum_{cI'K'} \Gamma(aIK,cI'K')u_{J\nu}(cI'K') \nonumber \\
&& +\sum_{cI'K'} \Delta(aIK,cI'K')v_{J\nu}(cI'K'), \label{eq:red2} \\
\Gamma(aIK,cI'K') &=& \frac{1}{2}\sum_{Lbd} F_{acdb}(L)q_{K'K}^{(L,0)}(db)
\sqrt{(2L+1)(2I+1)}  \nonumber \\
& \times& (-1)^{j_a +I +J} \left\{\begin{array}{ccc}j_a &j_c &L \\
I' & I & J \end{array}\right\} (IKLK'-K|I'K'), \label{eq:gam} \\
\Delta(aIK,cI'K') &=& \frac{1}{2}\sum_{Lbd} G_{acdb}(L)
\Delta_{K'K}^{(L,0)}(db)\sqrt{(2L+1)(2I+1)}  \nonumber \\
& \times& (-1)^{j_a +I +J} \left\{\begin{array}{ccc}j_a &j_c &L \\
I' & I & J \end{array}\right\} (IKLK'-K|I'K'). \label{eq:del} 
\end{eqnarray}
In Eqs.\ (\ref{eq:red1}) and (\ref{eq:red2}) we have set $\epsilon_a^{\prime}
=\epsilon_a^{\prime\prime}=\epsilon_a$.

We introduce a condensed notation for (\ref{eq:red1})
and (\ref{eq:red2}), writing them in the operator form
\begin{eqnarray}
{\cal E}_{J\nu}\Psi_{J\nu} &=& \hat{{\cal K}}\Psi_{J\nu} +\hat{\omega}
\Psi_{J\nu},    \label{eq:red3} \\
\hat{{\cal K}}& =& \left( \begin{array}{cc} \epsilon + \Gamma & \Delta \\
\Delta & -\epsilon -\Gamma \end{array} \right), \label{eq:K} \\
\hat{\omega} &=& \left( \begin{array}{cc} \omega & 0 \\ 0 & \omega 
\end{array} \right).
\end{eqnarray}

We solve these equations in the approximation that forms part 
of the definition of the core-particle model.  Again we consider first the 
simplest case where only the ground state band of the cores is included.
The extension to excited bands and interband coupling can be dealt with
in analogy to the treatment described for the intrinsic system.  
Let the physical 
solutions of (\ref{eq:red3}) with $\hat{\omega}=0$ be designated 
as $\Psi_{J\nu}^{(0)}$, with corresponding energies ${\cal E}_{J\nu}^{(0)}$.  
Here the symbol $\nu$ abbreviates the set $(\kappa\tau)$.
We approximate the solutions of the full equation by the expansion
\begin{equation}
\Psi_{J\nu} = \sum_{\nu'} {\cal C}_{\nu\nu'}^J\, \Psi_{J\nu'}^{(0)}. 
\label{eq:ex1}
\end{equation}
The introduction of this expansion into (\ref{eq:red3})  leads immediately 
to the standard eigenvalue problem
\begin{eqnarray}
{\cal E}_{J\nu}{\cal C}_{\nu\nu'}^J&=&{\cal E}_{J\nu'}^{(0)}{\cal C}_{\nu\nu'}^J
+\sum_{\nu''}{\cal U}^J_{\nu'\nu''}{\cal C}^J_{\nu\nu''},\label{eq:final2} \\
{\cal U}^J_{\nu'\nu''} &=&\tilde{\Psi}_{J\nu'}^{(0)}\hat{\omega}
\Psi_{J\nu''}^{(0)}.  \label{eq:final3}
\end{eqnarray}
This equation is to be compared with Eq.\ (\ref{eq:final}), to which it is
equivalent as long as $\hat{\omega}$ has the form assumed in the derivation 
of the latter.   In fact, (\ref{eq:final2}) has an advantage in the case 
that the excitation spectrum is not conveniently expressed in algebraic form,
but its numerical values are known from experiment.

We can extend the theory to include multiple bands in the core nuclei.
We use the labels $K\sigma$ to distinguish the different bands and now take
as a zeroth approximation the coupling of the odd particle to a single one of 
these bands.  The theory is, to start with, the same as that described 
above except
that we must distinguish the results for the various cores, and this is done
by a superscript $K\sigma$.  In so far as the multipole fields and pairing
fields for the band $K\sigma$ are almost equal to those for the ground band,
the energies ${\cal E}_{J\nu}^{K\sigma(0)}$ are almost independent of 
$K\sigma$.  We prefer to lift this degeneracy  by shifting each of these
energies by $E^{K\sigma}$, the band head energy, and redefining $\hat{\omega}$
to be the excitation energy above the band head in each case.  The step
that follows is to introduce the mixing due to the core excitations and again
only the change in notation already specified is necessary to record the 
equations that generalize (\ref{eq:final2}) and (\ref{eq:final3}).

The final step is to include the further mixing due to interband multipole
fields (assuming that such mixing for the pairing fields can be neglected).
For this purpose, we decompose $\hat{\Gamma}$ into an intraband piece
(subscript $d$) and an interband part (subscript $od$), the latter 
having so far been neglected, according to the equation
\begin{equation}
\hat{\Gamma}= \tau_3(\hat{\Gamma}_d +\hat{\Gamma}_{od}),  \label{eq:decom}
\end{equation}
where $\tau_3$ is the usual Pauli matrix.
The perturbation previously neglected is dealt with by the expansion
\begin{equation}
\Theta_{J\rho} = \sum_{\nu K\sigma}{\cal D}^J_{\rho,\nu K\sigma}
\Psi_{J\nu}^{K\sigma},    
\end{equation}
where the mixing coefficients are determined by the conditions
\begin{eqnarray}
{\cal E}_{J\rho}{\cal D}^J_{\rho,\nu K\sigma}&=& {\cal E}^{K\sigma}_{J\nu}
{\cal D}^J_{\rho,\nu K\sigma}  \nonumber \\
&&+ \sum_{\nu' K'\sigma'}{\cal G}^J_{\nu K\sigma,\nu' K'\sigma'}
{\cal D}^J_{\rho, \nu' K'\sigma'}, \\
{\cal G}^J_{\nu K\sigma,\nu' K'\sigma'}&=& \tilde{\Psi}^{K\sigma}_{J\nu}
\tau_3\hat{\Gamma}_{od}\Psi^{K'\sigma'}_{J\nu'}.
\end{eqnarray}

\section{Improved algorithm}

The main source of difficulty perceived in the solution of the KKDF
equations is that the 
set of solutions is over-complete by a factor of two.
 This is a consequence of the fact
that the basis states form an over-complete (and, consequently,
non-orthogonal set).  Thus half of the states found by solving
the EOM are not physical and have to be identified and removed. 
The technique previously used to perform this task has now been
understood to be unnecessarily complicated.  

In the previous approach \cite{pav:1,DF3}
the Hamiltonian is first decomposed into symmetric and
anti-symmetric parts with respect to particle-hole conjugation. If
only the anti-symmetric part is diagonalized, then for every positive
energy eigenvalue there is a negative partner. From the BCS theory we
know that the positive eigenvalues are the physical solutions and the
negative eigenvalues the non-physical ones. Then the symmetric part is
turned on ``slowly'' and at every step the physical solutions are
identified using a projection operator built from the wavefunctions of
the previous step. 
Since the equations of motion have to be solved at each step,
the time needed to perform the calculation is correspondingly longer
than for a single diagonalization.
(In most applications a typical number of steps is 5.)

A simpler and quicker approach has now been identified.  Since the problem
decomposes into subproblems involving states of a fixed angular momentum,
we can invoke the no-crossing theorem.  This means that the relative
order in energy of the physical and of the non-physical states does not
change as we turn on the symmetric part of the Hamiltonian.
If the lower half of the  states (negative in particular) are the
unphysical ones in the  BCS limit, 
then at the physical limit where the full Hamiltonian is used, 
the lower half of the states are again the unphysical ones.
Consequently, we need only to solve the equations of motion
at the two limits, the BCS limit and the full Hamiltonian limit.
These remarks about the technique of solution apply not only to
the strong coupling examples studied in the next section, but also
to less straightforward applications of the KKDF method.

\section{Applications}
\begin{figure}
\centerline{\epsfysize=8cm \epsffile{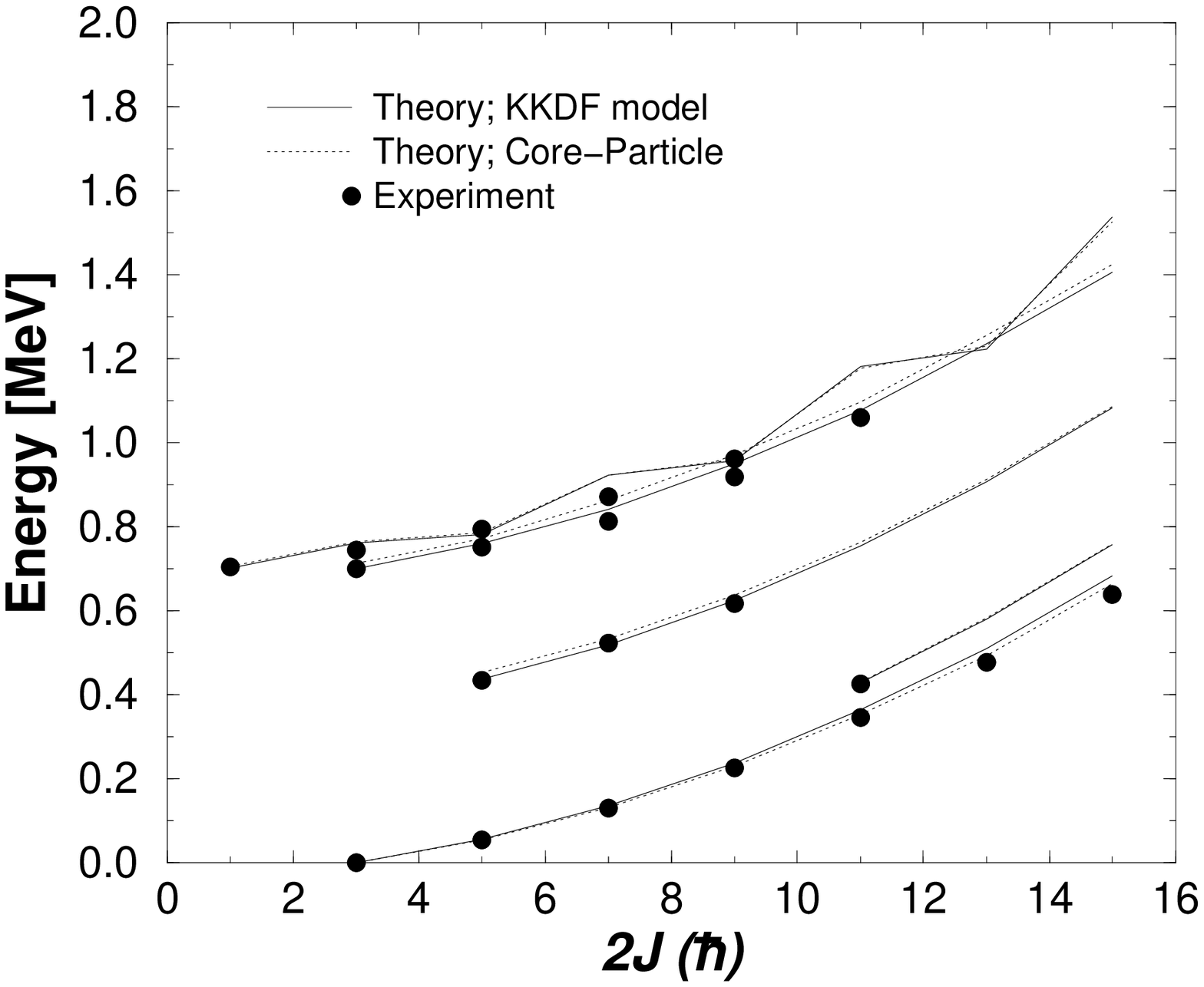}}
\caption{ \label{fig:Gd157_lev}
Negative parity energy levels for $^{157}$Gd. The circles 
correspond to the experimental values, the solid line 
to the KKDF model and the dotted line to the core-particle 
coupling model.}
\end{figure}

\begin{figure}
\centerline{\epsfysize=8cm \epsffile{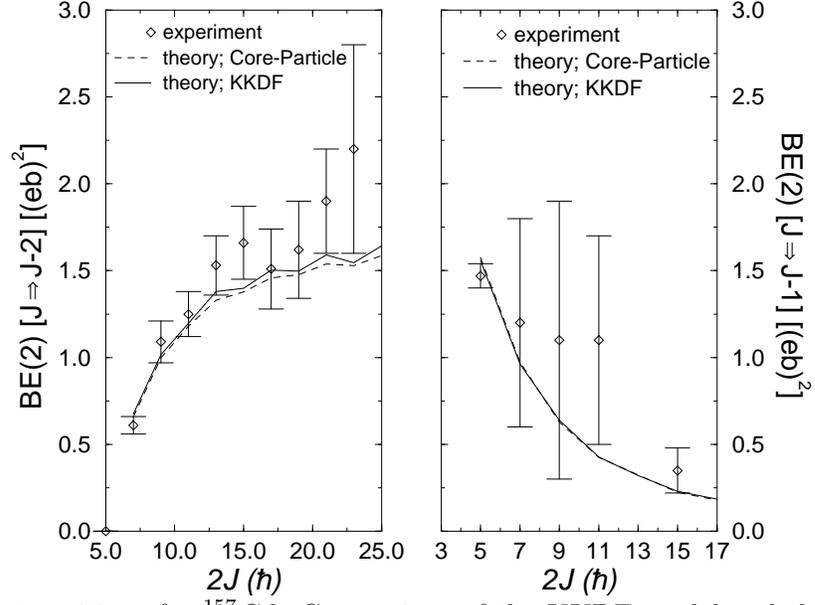}}
\caption{ \label{fig:Gd157_be2}
B(E2) transitions for $^{157}Gd$. 
Comparison of the KKDF model and the core-particle coupling model.
The points with error bars are the experimental data, the 
dashed lines result from the core-particle model and the 
solid lines from the KKDF model}
\end{figure}

We illustrate the remarks of the previous sections with applications
to a pair of well-deformed nuclei.
The first application is to the nucleus $^{157}$Gd, which we have
studied previously  \cite{pav:1,pav:2}. 
$^{157}$Gd is a well deformed nucleus and
thus suitable for application of the strong coupling core-particle model.  
To recall a few details, we used a large
single-particle space (including all states from 5 major shells).  The
energies and matrix elements of these single-particle levels were
calculated using the Woods-Saxon potential.  The odd neutron is 
coupled to the cores $^{156}$Gd and $^{158}$Gd, which are represented
not only by their ground bands, but also by several excited bands,
as was found necessary to fit all the observed bands of 
$^{157}$Gd.  The core excitation energies,
$\omega_{I}$, were given by phenomenological formulas tuned to experiment.
In the same way as in the previous papers, the strength of
the quadrupole field is treated as a free parameter and the values of
the single-particle energies found from Woods-Saxon calculations are
allowed to vary by $\pm 5 \%$. 
First we solved the EOM problem of the full KKDF 
model and fixed the strength of the quadrupole force and
 the single-particle energies in order to achieve the best fit.
Then we solved the EOM for the core-particle model as described in 
Sec.~\ref{cplab}, using the same parameters.  
The results are show in Fig.~(\ref{fig:Gd157_lev}). 
We can see from the figure that the two models give very similar 
results. In Fig.~(\ref{fig:Gd157_be2}) we show the result of the 
B(E2) calculations. Again it is clear that the two models give
very similar results.

\begin{figure}
\centerline{\epsfysize=8cm \epsffile{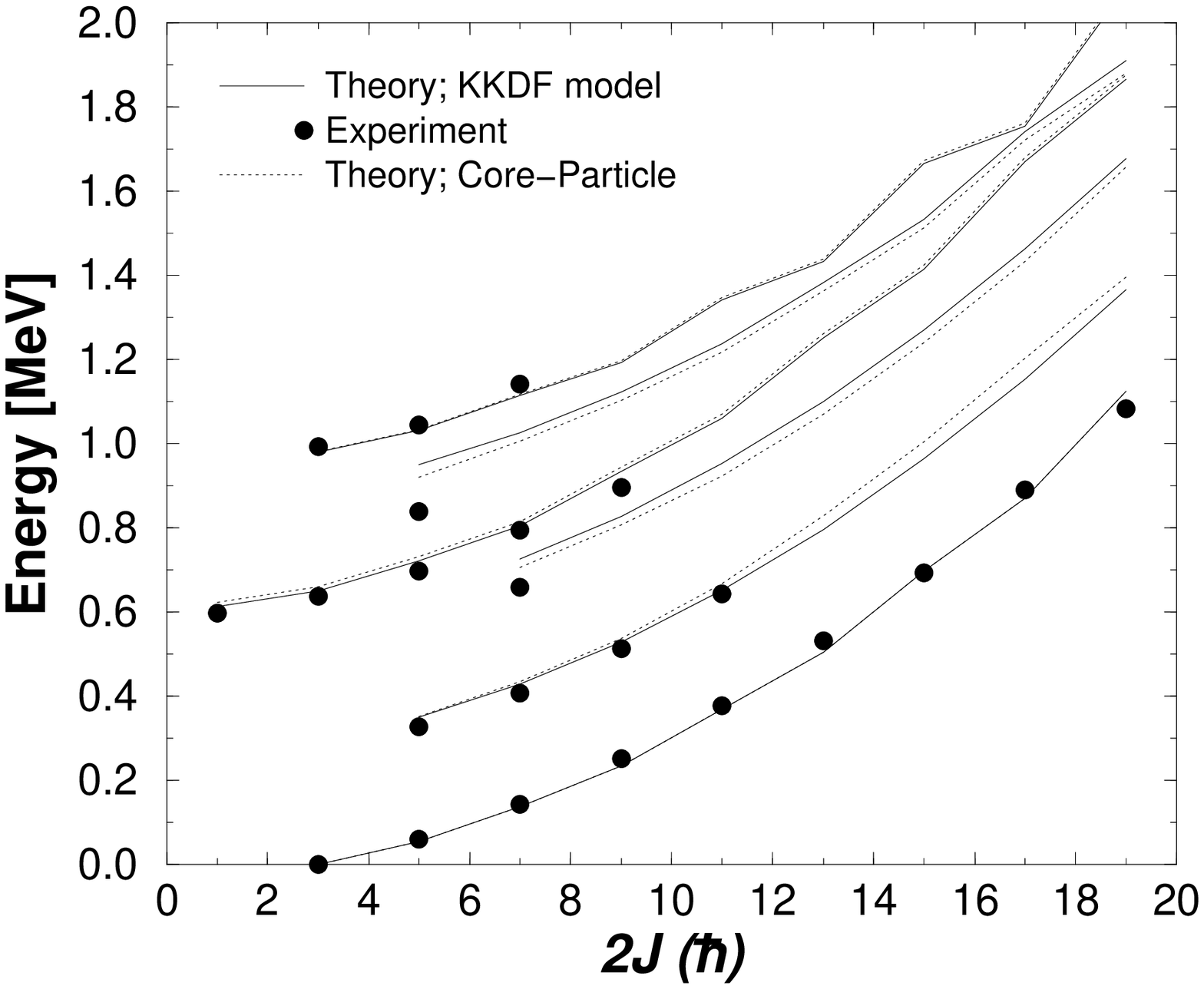}}
\caption{ \label{fig:Tb157_lev}
Positive parity energy levels for $^{157}$Tb. The circles 
correspond to the experimental values, the solid line 
to the KKDF model and the dotted line to the core-particle 
coupling model.}
\end{figure}

The second application was to the proton spectrum of 
 $^{157}$Tb, with $^{156}$Gd and $^{158}$Dy cores. 
We used the same method as described above 
and the results are shown in Fig.~(\ref{fig:Tb157_lev}). 
The conclusion is the same as in the previous application,
namely that the two methods give very similar results.
Observed B(E2) values are too few to allow a meaningful comparison.

To the extent that the examples chosen are typical,
it is apparent that 
for well-deformed nuclei the strong coupling
core-particle model gives almost as good results as the full KKDF model.
We emphasize, however, the greater range of validity of the KKDF model, in
particular to cases such as transitional nuclei \cite{pav:4,pav:thesis}, 
where none of the usual
traditional versions of the core-particle model is applicable.

\section{Discussion and concluding remarks}

In this paper, we have studied a semi-microscopic core-particle coupling
theory, the KKDF theory, and particularly its relationship to the 
traditional strong coupling core-particle model. The KKDF theory is 
formulated in the laboratory system of coordinates, and as such, can
be applied both to the spherical vibrational (weak coupling) and deformed
rotational (strong coupling) regimes, as well as to transitional cases.  
A significant portion of this paper has been devoted to transforming
the KKDF equations from the laboratory to the intrinsic system of coordinates,
the latter defined only for the well-deformed regime.  We have pointed out 
the additional approximation necessary to reduce the KKDF equations to 
those of the usual core-particle limit.  We have then applied both the 
full and the limiting model to a few illustrative nuclei and found only
small differences in the numerical results.  This justification is, for our
purposes, less significant than it would have been in the past, since we have
also formulated an improved algorithm that renders the KKDF equations 
essentially as simple to deal with as the defined approximation.

The reason for the good agreement between  the approximate and the complete
theory obviously expresses the fact that there is little mixing between
physical and unphysical states as we ``turn on'' the coupling that is 
initially suppressed in our approach.  This means that they stay well
separated in energy.  We can expect this situation to change for applications
where there are multiple avoided crossings.

\appendix
\renewcommand{\theequation}{\Alph{section}.\arabic{equation}}
\setcounter{equation}{0}

\section{Some details of the derivation of the core-particle coupling
model}

We provide some details of the derivations of Eqs.\ (\ref{eq:cp1})
and (\ref{eq:cp2}).  The first terms that require special attention 
are those involving the excitation energy in the even nuclei.  
We immediately do the sum over $M,\mu$.  Now consider Eq.\ 
(\ref{eq:cp1}), where we encounter the term 
\begin{equation}
(\omega_{IK}-E_K)(J\mu'j_a \kappa_a |IK) = (j_a -\kappa_a J\mu'|\frac{1}
{2{\cal I}_K}[ {\bf (J+j)^2} -K^2]|IK) ,      \label{eq:a2}
\end{equation}
(suppressing mass number).  We can replace the combination ${\bf (J+j)^2}$ by
\begin{equation}
J(J+1) +j_a(j_a +1) + 2(K-\kappa_a)\kappa_a +j_- J_+ +j_+ J_- .  
\label{eq:a3}    \end{equation}
Applying the standard algebra of the raising and lowering operators and
shifting the variables $\kappa_a$ as required for these terms, we thus
obtain additional contributions of single-particle type as well as the 
Coriolis coupling.  

We consider next the contributions of the multipole and pairing
fields, a calculation that requires most of the modest labor involved in    
the derivation of Eqs.\ (\ref{eq:cp1}) and (\ref{eq:cp2}).   As an example
of what is involved, we compute the contribution of the even multipoles
to the right hand side of (\ref{eq:cp2}), which we label $T(\Gamma^{\dag}:
JKj_a\kappa_a)$.  Utilizing Eqs.\ (\ref{eq:xyou}) and (\ref{eq:q}), 
we must evaluate the expression
\begin{eqnarray}
T(\Gamma^{\dag}:JKj_a\kappa_a)&=&\frac{1}{2}\sum(-1)^{j_c +m_a +\kappa_c 
-\kappa_a}\sqrt{\frac{(2j_a+1)(2I+1)}{(2j_c+1)(2I'+1)}}  \nonumber \\
&&\times F_{acdb}(L)q^{(L,0)}_{K'K}(db)  \nonumber\\
&&\times (j_a-m_a j_c m_c|LM_L)(IMLM_L|I'M')(I'M'Jm_c-M'|j_c m_c)\nonumber\\
&&\times(IMJm_a-M|j_a m_a)(JK-\kappa_a j_a\kappa_a|IK) \nonumber \\
&&\times(IKLK'_K|I'K')(JK'-\kappa_c j_c\kappa_c|I'K')\phi_{J,K'-\kappa_c}
(j_c\kappa_c).    \label{eq:TG}
\end{eqnarray}
In this equation the sum is over all angular momentum variables not indicated
explicitly on the left hand side except for $m_a$, which disappears from the
final result.

To evaluate this expression, we first study the partial sum
\begin{eqnarray}
S&=&\sum_{m_c} (-1)^{j_c+m_a}\sqrt{\frac{2j_a+1}{2j_c+1}} \nonumber \\
&&\times(j_a-m_a j_c m_c|Lm_c-m_a)(IMLm_c-m_a|I'M')(I'M'Jm_c-M'|j_c m_c)
\nonumber\\
&=&(-1)^{j_a+I'+J}\sqrt{(2L+1)(2I'+1)}  \nonumber \\
&&\times\left\{\begin{array}{ccc}j_a&j_c&L\\I'&I&J\end{array}\right\}
(IMJm_a-M|j_a m_a),
\label{eq:S}
\end{eqnarray}
which can be derived from Edmonds (6.2.7).  The sum over $M$ then removes
two more CG coefficients from (\ref{eq:TG})  The next step is to apply 
Edmonds (6.2.6) to evaluate the sum over $I'$,  
leading to a final trivial sum over $I$.  We thus find
\begin{eqnarray}
S'&=&\sum_{II'}(-1)^{j_a+I'+J}\sqrt{(2L+1)(2I+1)}
(JK-\kappa_a j_a\kappa_a|IK)\nonumber\\
&&\times(IKLK'-K|I'K')(JK'-\kappa_c j_c\kappa_c|I'K')
\left\{\begin{array}{ccc}j_a&j_c&L\\I'&I&J\end{array}\right\} \nonumber \\
&=& (-1)^{\kappa_c-\kappa_a+j_c+\kappa_c}(j_c-\kappa_c j_a\kappa_a|LK'-K).
\label{eq:S'}
\end{eqnarray}

Equations (\ref{eq:S}) and (\ref{eq:S'}) are the essential results 
for the evaluation of (\ref{eq:TG}) leading
to the appropriate term in (\ref{eq:cp2}).  The pairing term in the same 
equation and the multipole and pairing terms in (\ref{eq:cp1}) can be shown
(after straightforward transformations for the latter) to involve the
same basic sums, up to phase factors.

\setcounter{equation}{0}
\section{Contribution of odd multipole operators}

In the main text, we have included in the general core-particle 
equations (\ref{eq:cp1}) and (\ref{eq:cp2}) only contributions from
even electric multipole-multipole interactions, assuming that the included 
bands are all of the same parity.  If we wish to include
odd electric multipole-multipole forces to lowest order, we must replace the 
matrix element Eq.\ (\ref{eq:q}) by the value of the matrix element
\begin{eqnarray}
&&\langle I'M'K'|B^{\dag}_{LM_L}(db)|IMK\rangle  \nonumber \\ &&=
q_{K'K}^{(L,1)}(db)\sum_\lambda \langle I'M'K'|\{D^{L}_{M_L\,K'-K+\lambda},
I'_\lambda\}|IMK\rangle.   \label{eq:q1}
\end{eqnarray}
Here $I'_\lambda$, $\lambda=\pm1,3$ are the spherical tensor components of
the angular momentum in the ``intrinsic" system, expressed in terms of the 
Cartesian components (where they differ) by the equations
\begin{eqnarray}
I'_{+1} &=&-\frac{1}{\sqrt{2}}(I'_1 +iI'_2) =-\frac{1}{\sqrt{2}}I'_+
\nonumber \\
I'_{-1} &=&\frac{1}{\sqrt{2}}(I'_1 -iI'_2) =\frac{1}{\sqrt{2}}I'_- .
\label{eq:I'}
\end{eqnarray}
With the help of the well-known matrix elements of the intrinsic components,
Eq.\ (\ref{eq:q1}) takes the value
\begin{eqnarray}
&&\langle I'M'K'|B^{\dag}_{LM_L}(db)|IMK\rangle =
q_{K'K}^{(L,1)}(db) (IMLM_L|I'M')  \nonumber \\
&&\times \{-\frac{1}{\sqrt{2}}[(IK-1LK'-K+1|I'K')\sqrt{(I+K)(I-K+1)}\nonumber\\
&&+\sqrt{(I'-K')(I'+K'+1)}(IKLK'-K+1|I'K'+1)]  \nonumber \\
&&+\frac{1}{\sqrt{2}}[(IK+1LK'-K-1|I'K')\sqrt{(I-K)(I+K+1)}\nonumber \\
&&+\sqrt{(I'+K')(I'-K'+1)}(IKLK'-K-1|I'K'-1)]  \nonumber \\
&& +(K+K')(IKLK'-K|I'K')\}.    \label{eq:L1}
\end{eqnarray}

With these values, we are now in a position to evaluate the contributions
of an odd electric multipole force to our core-particle coupling equations.
We first consider the contributions to (\ref{eq:cp2}).  As an example,
consider the first term of (\ref{eq:L1}).  The calculation parallels 
that described in the previous appendix.  The sum (\ref{eq:S}) repeats itself
in every case.  The sum (\ref{eq:S'}) is replaced, in general, by different
expressions.  In the case of the first term of (\ref{eq:L1}),
 the sum $S'$ is replaced by the sum
\begin{eqnarray}
S'_1&=&\sum_{II'}(-1)^{j_a+I'+J}\sqrt{(2L+1)(2I+1)}
(JK-\kappa_a j_a\kappa_a|IK)\nonumber\\
&&\times(IK-1LK'-K+1|I'K')(JK'-\kappa_c j_c\kappa_c|I'K')  \nonumber \\
&&\times\left\{\begin{array}{ccc}j_a&j_c&L\\I'&I&J\end{array}\right\} 
\sqrt{(I+K)(I-K+1)}.    \label{eq:s1}
\end{eqnarray}
For $S'_1$, the sum over $I'$ can be carried out as before, 
but after this has been 
done, instead of a final normalization condition 
for CG coefficients, we encounter the sum 
\begin{eqnarray}
&&\sum_I (JK-\kappa_a j_a\kappa_a|IK)(JK-\kappa_a-1 j_a\kappa_a|IK-1)
\sqrt{(I+K)(I-K+1)}   \nonumber \\
&&=\sqrt{(J-K+\kappa_a+1)(J+K-\kappa)},  \label{eq:s2}
\end{eqnarray}
which involves the same ``trick" as used in the evaluation of the Coriolis
coupling.   Of the terms arising from (\ref{eq:L1}), the first and third
require the procedure just described, the second and fourth a similar
procedure in which we interchange the order of the sums on $I$ and $I'$,
and the fifth the same calculation as in the previous appendix.  We
also find that the first two terms are equal, as are the third and fourth.

Altogether, we find for the contribution to Eq.\ (\ref{eq:cp2}), the 
expression
\begin{eqnarray}
&&\sum_{bcd\kappa_c K'L}(-1)^{j_c +\kappa_c}F_{acdb}(L)q_{K'K}^{(L,1)}(db) 
\nonumber\\
&&\times[\frac{1}{\sqrt{2}}\sqrt{(J-K+\kappa_a+1)(J+K-\kappa_a)}
(j_c-\kappa_c j_a\kappa_a|LK-K'-1) \phi_{J,K-1-\kappa_a}(j_c\kappa_c) 
\nonumber \\
&& -\frac{1}{\sqrt{2}}\sqrt{(J+K-\kappa_a+1)(J-K+\kappa_a)}
(j_c-\kappa_c j_a\kappa_a|LK-K'+1) \phi_{J,K+1-\kappa_a}(j_c\kappa_c) 
\nonumber \\
&& -\frac{1}{2}(K+K')
(j_c-\kappa_c j_a\kappa_a|LK-K') \phi_{J,K-\kappa_a}(j_c\kappa_c). 
\label{eq:s3}
\end{eqnarray}
For conciseness of expression, we have not done the sum over $\kappa_c$.
In this form it can be shown that the corresponding contribution to 
Eq.\ (\ref{eq:cp1}) differs only by overall sign and by the replacement
\begin{equation}
(-1)^{j_c +\kappa_c} \rightarrow (-1)^{j_c +\kappa_a +L}.
\end{equation}

\end{document}